\documentclass[12pt]{article}

\usepackage{amsfonts}
\usepackage{bm}
\usepackage{cite}
\usepackage{mathrsfs}
\usepackage{amsmath}
\usepackage{graphicx}
\usepackage{hyperref}
\usepackage{verbatim}
\usepackage{color}
\usepackage[dvipsnames]{xcolor}
\usepackage{enumerate}
\usepackage{amsfonts}
\usepackage{amssymb}
\usepackage{authblk}
\usepackage[hang,flushmargin]{footmisc}
\usepackage{lipsum}
\usepackage[font=footnotesize]{caption}
\usepackage{soul}
\usepackage{empheq}
\usepackage{cancel}

\csname @addtoreset\endcsname{equation}{section}
\newcommand{\eal}[1]{\begin{equation} \begin{aligned} #1 \end{aligned}\end{equation}}
\newcommand{\lr}[1]{\langle #1 \rangle}

 \newcommand{\badat}{\begin{alignedat}}
 \newcommand{\eadat}{\end{alignedat}}
 
\def\bz{{\bar z}}

\def\be{\begin{eqnarray}}
\def\ee{\end{eqnarray}}
\def\beann{\begin{eqnarray*}}
\def\eeann{\end{eqnarray*}}
\def\beq{\begin{equation}}
\def\eeq{\end{equation}}
\def\ba{\begin{array}}
\def\ea{\end{array}}
\def\ben{\begin{enumerate}}
\def\een{\end{enumerate}}
\def\bea{\begin{eqnarray}}
\def\eea{\end{eqnarray}}

\def\5{\bar }
\def\6{\partial }
\def\7{\hat }
\def\4{\tilde }
\newcommand{\half}{\frac{1}{2}}

\def\cA{\mathcal{A}}
\def\cB{\mathcal{B}}

\def\cE{\mathcal{E}}
\def\cF{\mathcal{F}}
\def\cG{\mathcal{G}}

\def\cI{\mathcal{I}}

\def\cM{\mathcal{M}}
\def\cN{\mathcal{N}}
\def\cO{\mathcal{O}}
\def\cP{\mathcal{P}}

\def\cR{\mathcal{R}}
\def\cS{\mathcal{S}}

\hypersetup{final}

\begin{document}

 \begin{titlepage}
  \thispagestyle{empty}
  \begin{flushright}
  CPHT-RR054.082020 
  \end{flushright}
  \bigskip
  \begin{center}
	 \vskip2cm
  \baselineskip=13pt {\LARGE \scshape{Loops on the Celestial Sphere}}\\ 
   \vskip1.5cm 
   %\today
	% \vskip1.5cm
   \centerline{ 
   {Hern\'an A. Gonz\'alez}${}^a$, 
   {Andrea Puhm}${}^b$
   {and Francisco Rojas}${}^c$
   }

 \bigskip\bigskip
 
 \centerline{\em${}^a$ Facultad de Artes Liberales, Universidad Adolfo Ib\'a\~nez, Santiago, Chile}
 
\smallskip
 
\centerline{\em${}^b$   CPHT, CNRS, Ecole polytechnique, IP Paris, F-91128 Palaiseau, France}

\smallskip
 
\centerline{\em${}^c$   Facultad de Ingenier\'ia y Ciencias, Universidad Adolfo Ib\'a\~nez, Santiago, Chile}

\bigskip\bigskip
  
  \end{center}

\begin{abstract}
  \noindent
We study the effect of loop corrections to conformal correlators on the celestial sphere at null infinity. We first analyze finite one-loop celestial amplitudes in pure Yang-Mills theory and Einstein gravity. We then turn to our main focus: infrared divergent loop amplitudes in planar $\cN=4$ super Yang-Mills theory. We compute the celestial one-loop amplitude in dimensional regularization and show that it can be recast as an operator acting on the celestial tree-level amplitude. This extends to any loop order and the re-summation of all planar loops enables us to write down an expression for the all-loop celestial amplitude. Finally, we show that the exponentiated all-loop expression given by the BDS formula gets promoted on the celestial sphere  to an operator acting on the tree-level conformal correlation function, thus yielding, the \emph{celestial} BDS formula. 
\end{abstract}

 \end{titlepage}
\tableofcontents

% 
% \title{\vspace{-70pt} \Large{\sc Loops on the Celestial Sphere}\vspace{10pt}}
% \author[a]{\normalsize{Hern\'an A. Gonz\'alez}}
% \author[b]{\normalsize{Andrea Puhm}}
% \author[c]{\normalsize{Francisco Rojas}}
% 
% \affil[a]{{\small\textit{Facultad de Artes Liberales, Universidad Adolfo Ib\'a\~nez, Santiago, Chile}}}
% \affil[b]{{\small\textit{CPHT, CNRS, Ecole Polytechnique, IP Paris,F-91128 Palaiseau, France}}}
% \affil[c]{{\small\textit{Facultad de Ingenier\'ia y Ciencias, Universidad Adolfo Ib\'a\~nez, Santiago, Chile}}}
% 
% 
% \date{}
% \maketitle
% \thispagestyle{empty}
% \begin{abstract}
% 
% We study the effects of loop corrections to conformal correlators on the celestial sphere. We first analyze finite loop amplitudes for gluons and gravitons and we then consider infrared divergent amplitudes in planar $\cN=4$ super Yang-Mills theory. For the $\ell$-loop celestial case, we show that the correlator can be recast as an operator acting on the tree-level amplitude. We then focus the on the ressumation of all planar loops given by the BDS formula. We show that the exponential factor manifests on the celestial sphere again as an operator acting onto the tree-level conformal correlation function, thus yielding, the \emph{celestial} BDS formula.

% \end{titlepage}

% \newpage
% 
% \begin{small}
% {\addtolength{\parskip}{-2pt}
%  \tableofcontents}
% \end{small}
% \thispagestyle{empty}
% \newpage

\section{Introduction}

Celestial amplitudes reveal conformal properties of four-dimensional scattering amplitudes of massless particles as the standard plane wave basis is replaced by a basis of boost eigenstates. This is achieved by a Mellin transform applied to each external state in the scattering amplitude which maps plane waves labelled by the null momenta of the particles, or equivalently, their energy and a point on the two-sphere as well as their helicity to so-called conformal primary wavefunctions. The latter are labeled by the conformal dimension $\Delta$ and the spin $J$ under the two-dimensional global conformal group which arises from the action of the four-dimensional $SL(2,\mathbb{C})$ Lorentz group on the celestial sphere at null infinity.

Conformal wavefunctions have already been considered by Dirac~\cite{Dirac:1936fq}, but recent years have seen a surge in interest in part due to their role for a potentially holographic description of asymptotically flat spacetimes~\cite{deBoer:2003vf,Barnich:2009se,Barnich:2010eb,Strominger:2017zoo}. This has in part been driven by the realization that the asymptotic symmetry group of Einstein gravity at null infinity should include an extension of the BMS group that enhances the global conformal group on the celestial sphere to the full Virasoro symmetry group.\footnote{The Virasoro symmetry may be further enhanced to Diff($S^2$) - see~\cite{Donnay:2020guq} for a discussion of this point. See also~\cite{Donnay:2018neh,Banerjee:2018fgd,Stieberger:2018onx,Law:2019glh} for further results on symmetries of celestial amplitudes.}

To understand the properties of a putative holographically dual {\it celestial} CFT at null infinity of asymptotically Minkowski spacetimes a bottom-up approach has been pursued starting with the identification of a conformal basis of wavefunctions in~\cite{Pasterski:2017kqt,Donnay:2018neh,Law:2020tsg,Muck:2020wtx,Narayanan:2020amh} and the computation of celestial amplitudes for various spins~\cite{Cheung:2016iub,Pasterski:2017ylz,Cardona:2017keg,Schreiber:2017jsr,Stieberger:2018edy} which all take the form of two-dimensional correlation functions on the celestial sphere.
Furthermore, the celestial analogue of various soft theorems in quantum field theory have been obtained in~\cite{Nandan:2019jas,Pate:2019mfs,Adamo:2019ipt,Puhm:2019zbl,Fan:2020xjj,Banerjee:2020zlg}, while collinear limits of scattering amplitudes have been shown in~\cite{Fan:2019emx,Fotopoulos:2019tpe,Fotopoulos:2019vac,Pate:2019lpp,Banerjee:2020kaa} to extract from celestial amplitudes the operator product expansion of conformal primaries in the putative celestial CFT. The formalism for a relativistic partial wave expansion for celestial four-point amplitudes was developed in~\cite{Law:2020xcf} (see also~\cite{Nandan:2019jas}). A procedure for a celestial double copy relating celestial gravity and gauge theory amplitudes has been discussed in~\cite{Casali:2020vuy}. 

Most work so far has focused on celestial tree-level amplitudes while quantum effects have largely been neglected. One may thus wonder how the conformal structure uncovered at tree-level is affected once loop corrections are taken into account. 
The Mellin transform involves a sum over all energies that mixes the infrared and ultraviolet regimes. The main difficulty in computing celestial amplitudes at loop-level is thus due to the integration over internal momentum loops which render the Mellin integrals divergent.\footnote{These difficulties are circumvented when considering the celestial analogue of finite loop-amplitudes whose tree-level counter parts vanish. Such amplitudes were considered~in~\cite{Albayrak:2020saa} for gluons and gravitons and we will discuss the case of one-loop four-point amplitudes in pure Yang-Mills theory and Einstein gravity below.}
Nevertheless, there have been first attempts at understanding the conformal structure of flat space scattering amplitudes beyond tree level in~\cite{Banerjee:2017jeg} which studies celestial loop effects in massless scalar field theory with $\phi^4$ interaction.

Here we would like to initiate the study of divergent loop corrections to celestial tree-level amplitudes involving gluons as external massless states. Due to gauge invariance, the external states are guaranteed to reach null infinity at any order in perturbation theory. We will focus on the case of MHV four-point amplitudes in planar $\mathcal{N}=4$ Super Yang Mills theory. This has the benefit that the four-gluon amplitude is known to all orders in the loop expansion. 
To control infrared divergences we employ the dimensional regularization method known as four-dimensional helicity (FDH) scheme~\cite{Bern:2002zk}. The momenta of the internal particles are taken to be in $D=4-2\epsilon$ dimensions while the momenta of the external particles remain in $D=4$ as do the polarization vectors for both external and internal particles.
The four-gluon MHV amplitude in planar $\mathcal{N}=4$ SYM can be conveniently written in factorized form in terms of the tree-level amplitude containing the helicity structure and an infrared divergent piece containing the information about the loop-order. 

Our main result is that this infrared divergent factor gets promoted, after the Mellin transform from the momentum basis to the conformal basis, to a differential operator.
Thus the celestial one-loop amplitude in planar $\mathcal{N}=4$ SYM can be recast as an operator acting on the celestial tree-level amplitude.\footnote{Similar conclusions have been recently hinted at for QED and gravity~\cite{Stromingerstring}.} This structure is shown to persist at any loop-order. The effect of this {\it celestial loop operator} is to simultaneously shift the conformal dimensions of all the external particles. The operator is dressed with a loop-order dependent function of the conformal cross ratio and diverges as the dimensional regulator is taken to zero. 

Given that the Mellin transform effectively replaces the notion of energy with that of conformal scaling dimension it is not too surprising that loop effects should manifest themselves as a change in the latter. Interestingly enough, though, celestial loop effects can be recast as an operator statement. This also resonates with the recent presentation  in~\cite{Casali:2020vuy} of tree-level graviton and gluon amplitudes as operators acting on scalar amplitudes when the asymptotic states are taken to be in the conformal basis.

In momentum-space the four-gluon amplitude is known to all orders in the loop expansion. Moreover, the re-summation of all loop contributions exponentiates yielding the BDS formula found by Bern, Dixon and Smirnov~\cite{Bern:2005iz} based on an iterative relation uncovered by Anastasiou, Bern, Dixon and Kosower~\cite{Anastasiou:2003kj}\footnote{This was later confirmed in the strong coupling regime by Alday and Maldacena \cite{Alday:2007hr}.}. 
Here we provide its celestial analogue. At any loop-order there exists an operator acting on the celestial tree-level amplitude which results in a loop-order dependent shift in the conformal dimensions. The re-summation of all celestial loop operators exponentiates, thus yielding the celestial BDS formula.

While we have focused here on the four-gluon amplitude one could analyze higher-point amplitudes at loop-level in $\mathcal{N}=4$ SYM as well as other theories with less symmetry. A preliminary study shows that a similar operator structure as the one we uncovered here seems to arise for pure Yang-Mills theory~\cite{GRwip}. It would be interesting to investigate whether this pattern persists in other theories.
In~\cite{Stieberger:2018edy} the authors computed celestial tree-level amplitudes of four massless states in the open sector of the type I string. A natural next step is to compute the celestial one-loop string amplitude and see if it can be recast as an operator acting on the celestial tree-level string amplitude. If that is the case, and since conformal invariance on the worldsheet intimately ties infrared divergences to ultraviolet ones through the open/closed string duality, it would be appealing to explore how this duality manifests itself on the celestial sphere. 
We leave these interesting questions for the future.

This paper is organized as follows. We begin with a review in section~\ref{sec:Setup} where we set up the necessary notation. A convenient basis for discussing celestial amplitudes at tree and loop-level is introduced in section~\ref{ssec:HelicityBasis}. In section~\ref{ssec:CelestialAmplitudes} we review some of the salient features of celestial amplitudes and give a tree-level example in section~\ref{ssec:TreeExample}. We discuss loop-amplitudes in section~\ref{sec:Loops}. First we compute in section~\ref{ssec:Finite1LoopPureYM} celestial amplitudes for one-loop processes in Yang-Mills theory for which the tree-level amplitudes vanish. We then move on to the main focus of this paper in section~\ref{ssec:DivergentLoops} and analyze how divergent loop corrections in planar $\mathcal{N}=4$ Super Yang-Mills theory correct the corresponding tree-level celestial correlators. We show in section~\ref{ssec:Divergent1LoopSYM} that the one-loop four-gluon celestial amplitudes can be recast as an operator acting on the celestial tree-level amplitude and generalize this result to $\ell$ loops in section~\ref{ssec:AllLoopSYM}.
The exponentiated re-summation of the all-loop result then gets promoted to an operator statement yielding the celestial analogue of the BDS formula. In appendix~\ref{app:MoreCelestial} we make use of the general expression for celestial amplitudes introduced in section~\ref{sec:Setup} to analyze tree-level amplitudes in massless QED in appendix~\ref{app:MoreCelestialQED} and finite one-loop amplitudes in Einstein gravity in appendix~\ref{app:MoreCelestialgravity}.

\section{Setup}
\label{sec:Setup}
In this section we lay out a convenient basis for expressing celestial amplitudes at tree and loop-level.
We consider scattering amplitudes in four dimensions\footnote{We work in $(+---)$ signature.}  with external massless particles. 
Each external particle is labelled by a momentum $p_i^\mu$, a helicity $\ell_i$, and a sign distinguishing incoming from outgoing states. We parameterize $p_i^\mu$ in terms of points $(z_i,\bar{z}_i)$ on the two-dimensional celestial sphere through the map
\be
\label{p}
 p_i^\mu=\frac{1}{2} \omega_i \left(1+|z_i|^2, z_i+\bar{z}_i,-i(z_i-\bar{z}_i),1-|z_i|^2 \right)\,,
\ee
with $\omega_i\geq0$.

\subsection{A convenient basis for 4-point amplitudes}
\label{ssec:HelicityBasis}

Our focus will be on four-point amplitudes
\begin{equation}
\label{eq:4ptscattering}
 \cA\left(\{\omega_i,\ell_i,z_i,\bar{z}_i\}\right)=A\left(1^{\ell_1},2^{\ell_2},3^{\ell_3},4^{\ell_4}\right)   \delta(p_1+p_2-p_3-p_4) \,.
\end{equation}
Without loss of generality, we assume that the stripped amplitude can be split as 
\be
\label{gen4pt}
A\left(1^{\ell_1},2^{\ell_2},3^{\ell_3},4^{\ell_4}\right)=  \cB(s,t) \; \cR_{(\ell_1,\ell_2,\ell_3,\ell_4)}\,,
\ee
where $\cB(s,t)$ is a (not necessarily analytic) function of two of the Mandelstam variables $s=(p_1+p_2)^2$, $t=(p_1-p_4)^2$ and $u=(p_1-p_3)^2$, and the rational function $\cR_{(\ell_1,\ell_2,\ell_3,\ell_4)}$ carries the helicity structure of the external particles. The latter is a function of the spinor products $\langle i j\rangle$ and $[i j]$ defined by\footnote{The spinors $\lambda^{\alpha}_{i}$ and $\tilde{\lambda}^{\dot \alpha}_{i}$ can be read off from
\be
p^{\dot \alpha \alpha}=\bar{\sigma}^{\dot \alpha \alpha}_{\mu} p^{\mu} = \omega 
\begin{bmatrix}
    1 & z  \\
    \bar{z} & |z|^2 
  \end{bmatrix} =\underbrace{\sqrt{\omega} \begin{bmatrix}
    1   \\
    \bar{z} 
  \end{bmatrix} }_{\tilde{\lambda}^{\dot \alpha }}
 \underbrace{\sqrt{\omega} \begin{bmatrix}
    1 & z 
  \end{bmatrix} }_{\lambda^{\alpha } }.
  \nonumber
\ee
} 
\be
\label{prod}
\langle i j \rangle= \epsilon_{\alpha \beta} \lambda^{\alpha}_{i} \lambda^{\beta}_{j}= \sqrt{\omega_i \omega_j} z_{ij}\,,\quad 
[ i j ]= - \epsilon_{\dot \alpha \dot \beta} \tilde{\lambda}^{\dot \alpha}_{i} \tilde{\lambda}^{\dot \beta}_{j}= - \sqrt{\omega_i \omega_j} \bar{z}_{ij}\,,
\ee
where $z_{ij}=z_i-z_j$ and $\bar{z}_{ij}=\bar{z}_i-\bar{z}_j$, as well as the conformally invariant cross-ratio
\be
\label{eq:crossratio}
r=\frac{z_{12} z_{34}}{z_{23} z_{41}}\,,
\ee 
which is related to the four-dimensional scattering angle in the center of mass frame $\theta$ through 
\eal{\label{eq:rst}
r=-\frac{s}{t} = \csc^2\left(\frac{\theta}{2}\right)\,.
}
It follows that $r>1$.

Since all the information about the helicities of the external states is encoded in the function $\cR_{(\ell_1,\ell_2,\ell_3,\ell_4)}$, it must satisfy 
\be \label{hEVeq}
\hat{ \ell}_i \cR_{(\ell_1,\ell_2,\ell_3,\ell_4)} = \ell_i \cR_{(\ell_1,\ell_2,\ell_3,\ell_4)}\,,
\ee 
for the helicity operator for the $i$-th particle
\eal{
\hat{ \ell}_i = \frac{1}{2}\left(-\lambda^{\alpha}_i \frac{\partial}{\partial \lambda^{\alpha}_i } + \tilde{\lambda}^{\dot{\alpha}}_i \frac{\partial}{\partial \tilde{\lambda}^{\dot{\alpha}}_i } \right)\,.
}
We may express a solution to \eqref{hEVeq}  in terms of powers of Lorentz invariant functions $R_i$, with the defining property that $\hat{ \ell}_i R_j=\delta_{ij} R_j $, as
\be
\cR_{(\ell_1,\ell_2,\ell_3,\ell_4)}= r^{\alpha_1} (r-1)^{\alpha_2}  \prod^{4}_{i=1}R_i^{\ell_i}\,,
\ee
with $\alpha_1$ and $\alpha_2$ real numbers. 
A basis of functions $R_i$ that depend on the energies $\omega_i$ of the external particles was given in~\cite{Badger:2016uuq}. For our purposes it is more convenient to express $\cR$ in terms of functions that depend exclusively on the differences of points $(z_{ij}, \bar{z}_{ij})$ on the celestial sphere. Such a set of $R_i$ is given by
\begin{align}
\begin{split}
R_1=\left(\tfrac{[12][13] \langle23\rangle}{\langle 12\rangle \langle13 \rangle  [32]}\right)^{\frac{1}{2}}\,,&\quad R_2=\left(\tfrac{[12][23] \langle13\rangle}{\langle 12\rangle \langle2 3 \rangle  [31]}\right)^{\frac{1}{2}}\,,\\
R_3= \left(\tfrac{[13][23] \langle12\rangle}{\langle 13\rangle \langle 23 \rangle  [21]}\right)^{\frac{1}{2}}\,,&\quad R_4= \tfrac{[24]}{\langle 24 \rangle} \left(\tfrac{[31] \langle12\rangle \langle23\rangle}{ \langle 13\rangle  [12] [23]}\right)^{\frac{1}{2}}\,.
\end{split}
\end{align}
The four point amplitude \eqref{gen4pt} can then be expressed as 
\begin{multline}
\label{eq:4point}
A\left(1^{\ell_1},2^{\ell_2},3^{\ell_3},4^{\ell_4}\right)=r^{\alpha_1} (r-1)^{\alpha_2} \cB(s,t) \left(\tfrac{ z_{12} }{\bar{z}_{12} }\right)^{-\tfrac{1}{2}(\ell_1+\ell_2-\ell_3-\ell_4)} \left(\tfrac{ z_{13} }{\bar{z}_{13} }\right)^{-\tfrac{1}{2}(\ell_1-\ell_2+\ell_3+\ell_4)}\\ \times \left(\tfrac{ z_{23} }{\bar{z}_{23} }\right)^{\tfrac{1}{2}(\ell_1-\ell_2-\ell_3+\ell_4)}\left(\tfrac{ z_{24} }{\bar{z}_{24} }\right)^{-\ell_4} \,.
\end{multline}
Notice that no mention to perturbation theory has been made, and as long as the amplitude is of the form~\eqref{eq:4point}, the statements made so far apply to the exact four-point $\cS$-matrix element.

\subsection{Celestial amplitudes}\label{ssec:CelestialAmplitudes}
Celestial amplitudes are obtained by performing a Mellin transform on each of the $n$ external particles in a scattering process 
\be
\label{eq:4}
\tilde{\cA}\left(\{\Delta_i,J_i,z_i,\bar{z}_i\}\right)= \prod^n_{k=1} \left( \int^\infty_0  d\omega_k \omega^{\Delta_k-1}_{k}\right) \cA\left(\{\omega_i,\ell_i,z_i,\bar{z}_i\}\right)\,.
\ee
Under $SL(2,\mathbb{C})$ Lorentz transformations they have been shown to transform as~\cite{Pasterski:2017ylz,Pasterski:2017kqt,Stieberger:2018edy}
\begin{equation}
  \widetilde{\cA}\Big(\Big\{\Delta_j, J_i;\frac{a z_i+b}{cz_i+d},\frac{ \bar{a}  \bz_i+ \bar{b}}{ \bar{c} \bz_i+ \bar{d}}\Big\}\Big) =\prod_{j=1}^n \Big( (c z_j+d)^{\Delta_j+J_j} (\bar{c} \bz_j+\bar{d})^{\Delta_j-J_j}\Big) \widetilde{\cA}(\{\Delta_i,J_i;z_i,\bz_i\})\,,
\end{equation}
where $\Delta_j$ are the conformal dimensions and $J_j \equiv \ell_j$ the spins of operators inserted at the points $(z_j,\bz_j)\in S^2$. 
Celestial amplitudes thus share conformal properties with correlation functions on the celestial sphere. Indeed, evaluating \eqref{eq:4} for the four-point amplitude~\eqref{eq:4ptscattering} using 
\begin{multline}
\label{delta}
 \delta^{(4)}(p_1+p_2-p_3-p_4)=\frac{4}{\omega_4 |z_{14}|^2 |z_{23}|^2} \delta(r-\bar{r}) 
\\
  \times \delta\left(\omega_1 - \frac{z_{24}\bz_{34}}{z_{12}\bz_{13}}\omega_4\right)\delta\left(\omega_2 + \frac{z_{14}\bz_{34}}{z_{12}\bz_{23}}\omega_4\right)
 \delta\left(\omega_3 + \frac{z_{24}\bz_{14}}{z_{23}\bz_{13}}\omega_4\right)\,,
\end{multline}
yields the celestial four-point amplitude (up to a numerical factor) 
\be
\label{eq:Celestial4pt}
\tilde{\cA}\left(\{\Delta_i,J_i,z_i,\bar{z}_i\}\right)= f(r,\bar{r}) \prod_{i<j}^4 z_{ij}^{\frac{h}{3}-h_i-h_j}\bar{z}_{ij}^{\frac{\bar{h}}{3}-\bar{h}_i-\bar{h}_j}\,.
\ee
The conformally invariant expression $f(r,\bar{r})$ is given by
\be
\label{eq:f}
f(r,\bar{r}) =2  \delta(r-\bar{r}) \Theta(r-1) r^{\alpha_1+\frac{\Delta}{6}} (r-1)^{\alpha_2+\frac{\Delta}{6}} \int^{\infty}_0 dw\, w^{\tfrac{\Delta-6}{2} } \cB(r w, -w)\,,
\ee
with $\Delta \equiv \sum_{i=1}^4 \Delta_i$.
The conformal weights of the external wavefunctions introduced in~\eqref{eq:Celestial4pt} are given by 
\be
\label{eq:hs}
h_k=\frac{1}{2}(\Delta_k + J_k)\,, \quad \bar{h}_{k}=\frac{1}{2}(\Delta_k - J_k)\,,
\ee
where the two-dimensional spins $J_k$ are identified with the helicities $\ell_k$ of the four-dimensional particles crossing null infinity and the conformal dimensions $\Delta_k$ of finite energy wavefunctions are restricted to lie on the principal continuous series of the $SL(2,\mathbb{C})$ Lorentz group\cite{Pasterski:2017kqt}, namely 
\begin{equation}
 \Delta_k=1+i\lambda_k \; \text{with} \; \lambda_k\in \mathbb{R}\,, \quad  J_k\equiv \ell_k\,.
\end{equation}

\subsection{Example: celestial gluons at tree-level}
\label{ssec:TreeExample}

We illustrate the above prescription for celestial four-gluon amplitudes at tree-level in pure Yang-Mills theory which were first discussed in~\cite{Pasterski:2017ylz,Stieberger:2018edy}. 
The MHV four-gluon amplitude is given by~\eqref{eq:4ptscattering} with the stripped amplitude
\eal{\label{Atree}
A_{\rm tree}(1^-,2^-,3^+,4^+)&= g^2 \frac{\langle 12 \rangle^3}{\lr{23}\lr{34}\lr{41}}= g^2 \, r \frac{z_{12}\bar{z}_{34}}{\bar{z}_{12} z_{34}}\,,
}
where in the last step we used the constraint~\eqref{delta}.
Comparing with~\eqref{eq:4point} for $(\ell_1,\ell_2,\ell_3,\ell_4)=(-1,-1,1,1)$ we read off $\alpha_1=1$, $\alpha_2=0$ and the function $\cB(s,t)= g^2$ just corresponds to the coupling.\footnote{Here and henceforth we omit a factor $i(2\pi)^4$.} The corresponding celestial amplitude is given by the two-dimensional four-point correlation function~\eqref{eq:Celestial4pt} of gluons with conformal weights $(h_k,\bar{h}_k)=(\frac{i}{2} \lambda_k,1+\frac{i}{2}\lambda_k)$ for negative helicity and  $(h_k,\bar{h}_k)=(1+\frac{i}{2} \lambda_k,\frac{i}{2}\lambda_k)$ for positive helicity. The conformally invariant expression $f(r,\bar r)$ is given by
\eal{\label{eq:f_4gluontree}
f_{\rm tree}(r,\bar{r}) =2 g^2  \delta(r-\bar{r}) \Theta(r-1)r^{1+\frac{\Delta}{6}}(r-1)^{\frac{\Delta}{6}} \mathcal{I}(\lambda)\,,
}
where
\begin{equation}\label{eq:deltalambda}
 \mathcal{I}(\lambda)\equiv \int_0^\infty \frac{dw}{w}\, w^{i\frac{\lambda}{2}} =4\pi \delta(\lambda)\,,
\end{equation}
with $\lambda=\sum_{k=1}^4 \lambda_k$. Enforcing~\eqref{eq:deltalambda} sets $\Delta=4$ and thus yields
\begin{equation}
 f_{\rm tree}(r,\bar{r}) =8\pi g^2  \delta(r-\bar{r}) \Theta(r-1)r^{\frac{5}{3}}(r-1)^{\frac{2}{3}} \delta(\lambda)\,.
\end{equation}
Because the integral in~\eqref{eq:f} is marginally convergent for celestial tree-level Yang-Mills amplitudes it can be interpreted as a distribution given by~\eqref{eq:deltalambda}. This is no longer true for celestial amplitudes at loop-level which we will discuss in the next section. Before doing so let us comment that
while we have focused here on tree-level amplitudes in pure Yang-Mills theory in order to set the stage for the discussion of celestial gluon amplitudes at loop-level, the prescription discussed in sections~\ref{ssec:HelicityBasis}-\ref{ssec:CelestialAmplitudes} is more broadly applicable. 
In appendix~\ref{app:MoreCelestial} we analyze tree-level amplitudes in massless QED theory and finite one-loop Einstein gravity processes. 

\section{Celestial loop amplitudes}
\label{sec:Loops}
In this section we discuss how loop corrections modify celestial amplitudes. We first consider in section~\ref{ssec:Finite1LoopPureYM} processes in pure Yang-Mills theory that vanish at tree-level and yield finite one-loop amplitudes. In section~\ref{ssec:DivergentLoops} we move on to the more interesting case of MHV amplitudes in planar $\mathcal{N}=4$ super Yang-Mills theory. We compute the one-loop celestial amplitude in section~\ref{ssec:Divergent1LoopSYM} and extend this result to all loops in section~\ref{ssec:AllLoopSYM}.

\subsection{Finite one-loop amplitudes in Yang-Mills}
\label{ssec:Finite1LoopPureYM}

We consider loop corrected amplitudes for theories involving only external gluons. The simplest processes are the ones that vanish at tree-level.\footnote{
While this article was being prepared, we learned about the work of Albayrak, Chowdhury, and Kharel \cite{Albayrak:2020saa} which has some overlap with this subsection and with appendix~\ref{app:MoreCelestialgravity}.} This is the case for gluon amplitudes whose helicities  are $++++$ and $-+++$ as well as the opposite helicity cases. For the all helicity plus and all helicity minus amplitudes we have \cite{Bern:1991aq,Bern:1992ad,Bern:1993qk}  
\bea
\label{eq:12}
A(1^\pm,2^\pm,3^\pm,4^\pm)= g^4\left(\frac{[1 2][3 4]}{\langle 1 2 \rangle \langle 3 4 \rangle }\right)^{\pm 1}\,=  g^4 \left( \frac{\bar{z}_{12} \bar{z}_{34}}{z_{12} z_{34}}\right)^{\pm 1}\,,
\eea
while the mixed helicity amplitudes are given by
\be
\label{eq:14}
A(1^-,2^+,3^+,4^+)= g^4 \frac{\langle1 3\rangle [3 1][2 4]^2}{[1 2]  \langle 2 3 \rangle \langle 3 4 \rangle [41]}=  g^4 \frac{z_{13}  \bar{z}_{31}  \bar{z}^2_{24} } { \bar{z}_{12}  z_{23}   \bar{z}_{34}  \bz_{41}}\,,
\ee
with the opposite helicity amplitude obtained by exchanging $z_{ij} \leftrightarrow \bar z_{ij}$. In pure Yang Mills theory, these amplitudes correspond to the single trace and color ordered contribution to four-gluon processes. They also represent the leading contribution to photon-photon or gluon-gluon scattering from the box diagram of massless QED or QCD with a fermion running inside the loop \cite{Mahlon:1993si,Mahlon:1993fe,Bern:1995db}. 

Because the amplitudes \eqref{eq:12} and \eqref{eq:14} are again of the form \eqref{gen4pt}, the conformal weights are directly obtained from \eqref{eq:hs}. The new information from these loop processes in the bulk lies in the computation of the conformally invariant factor~\eqref{eq:f} which yields
\be
f_{\rm 1-loop}(r,\bar{r})= 8 \pi g^4 \delta (\lambda)  \delta (r-\bar{r})\Theta(r-1)(r-1)^{\frac{2}{3}} \times  \left\{
  \begin{array}{@{}ll@{}}
    r^{\frac{2}{3}} , & (\pm,\pm,\pm,\pm)  \\
    r^{\frac{5}{3}} , & (\mp,\pm,\pm,\pm)
  \end{array}\right.\,.
\ee 
Notice that here we were again able to make use of~\eqref{eq:deltalambda} reflecting the fact that the above amplitudes are finite even at one-loop. For infrared divergent amplitudes which we will discuss in the following the distribution involving the sum of the conformal dimensions will be promoted to an operator statement.

\subsection{Infrared divergent amplitudes in planar \texorpdfstring{$\mathcal{N}=4$}{N=4} SYM}
\label{ssec:DivergentLoops} 

We now turn to the more interesting case of MHV amplitudes and how loop corrections modify the celestial tree-level correlators involving gluons as external states. Because these amplitudes do not vanish at tree-level, their loop corrections (in momentum-space) typically suffer from both UV and IR divergences. While the former may be taken care of by renormalization, the latter are more subtle.\footnote{For a recent discussion of IR divergences see~\cite{Hannesdottir:2019rqq,Hannesdottir:2019opa}.} Here, to control infrared divergences we employ the dimensional regularization method known as four-dimensional helicity (FDH) scheme~\cite{Bern:2002zk}. The momenta of the internal particles are in $D=4-2\epsilon$ dimensions whereas all polarization vectors (internal and external) and the momenta of the external states remain in $D=4$. We focus on planar amplitudes in $\mathcal{N}=4$ Super Yang-Mills theory. Besides being UV finite, this has the benefit that the four-gluon amplitude is known to all orders in the loop expansion (in the 't~Hooft coupling).
The MHV four-gluon amplitude can be conveniently written as\footnote{See, for instance, section 4.3 in \cite{Henn:2014yza}.}
\be
A_{\rm all\,\,loops}= M_\epsilon A_{\rm tree}  \,,
\ee
where the scalar factor
\be
\label{eq:M-sum}
M_\epsilon = 1+ \sum_{\ell=1}^{\infty} a^\ell M^{(\ell)}_\epsilon\,, 
\ee
depends on the Mandelstam invariants $s$ and $t$ as well as the dimensional regulator. Here \mbox{$a\equiv \frac{g^2 N}{8\pi^2}(4\pi e^{-\gamma_E})^{\epsilon}$} is the 't~Hooft coupling and $\gamma_E$ is the Euler-Mascheroni constant. The tree-level amplitude $A_{\rm tree}$ is the same as in pure YM which for the choice of helicities $(1^-,2^-,3^+,4^+)$ is given in~\eqref{Atree}. 
As shown in~\cite{Bern:2005iz} the amplitude~\eqref{eq:M-sum} exponentiates according to the BDS formula
\begin{equation}
M_\epsilon={\rm exp}\left(\sum_{\ell=1}^\infty a^\ell (f^{(\ell)}_\epsilon M^{(1)}_{\ell \epsilon} +C^{(\ell)}+E^{(\ell)}_\epsilon)\right)\,,
\end{equation}
where $f^{(\ell)}_\epsilon$ are regular functions of $\epsilon$ and are directly related to the cusp and collinear anomalous dimensions, $C^{(\ell)}$ are numerical constants and $E^{(\ell)}_\epsilon$ is of order $\cO(\epsilon)$. In the following we will discuss the celestial analogue of the above statements.

\subsubsection{Celestial gluons at one-loop}
\label{ssec:Divergent1LoopSYM}

The one-loop contribution~\cite{Green:1982sw}
\begin{equation}
\label{eq:A-1loop}
A_{\rm 1-loop}=a M^{(1)}_\epsilon A_{\rm tree}   \,,
\end{equation}
is obtained by solving the scalar box integral 
\be
\label{eq:I4}
M^{(1)}_\epsilon = -\frac{1}{2} (\mu^2 e^{\gamma_E})^\epsilon \int \frac{d^Dp}{i\pi^{D/2}}\frac{st}{p^2(p+p_1)^2(p+p_1+p_2)^2(p+p_4)^2}\,,
\ee
where $\mu$ is the dimensional regularization scale. This integral can be explicitly evaluated in terms of hypergeometric functions (see for instance appendix E of \cite{Schubert:2001he})
\begin{multline}
\label{eq:I4-hyper}
M^{(1)}_\epsilon= -\frac{1}{\epsilon^2} e^{\epsilon \gamma_E}\gamma_{\Gamma}\frac{t}{\mu^2} \left[\left(\frac{\mu^2}{-t}\right)^{1+\epsilon} \hspace{-10pt}{}_2 F_1 \left(-\epsilon,1;1\!-\!\epsilon;1\!+\!\frac{t}{s}\right) \right. \\\left.- \left(\frac{\mu^2}{-s}\right)^{1+\epsilon} \hspace{-10pt} {}_2 F_1 \left(1,1;1-\epsilon;1\!+\frac{t}{s}\right)  \right] \,,
\end{multline}
where $\gamma_{\Gamma}=\Gamma(1+\epsilon)\Gamma^2(1-\epsilon)/\Gamma(1-2\epsilon)$. Note that we can express~\eqref{eq:I4-hyper} as
\be
M^{(1)}_\epsilon= \left(\frac{\mu^2}{-t}\right)^{\epsilon}\cF_1(r,\epsilon)\,,
\ee
where $\cF_1(r,\epsilon)$ depends on the Mandelstam invariants $s$ and $t$ only through the conformally invariant cross ratio $r=-s/t$ introduced in~\eqref{eq:rst} and has double poles in $\epsilon$ inherited from the dimensionally regularized one-loop integral whose explicit form can be read off from \eqref{eq:I4-hyper}. 

In computing the celestial four-gluon amplitude~\eqref{eq:Celestial4pt} at one-loop in planar $\mathcal{N}=4$ SYM we thus see that the main difference to the tree-level result lies in the (now divergent) Mellin integral in~\eqref{eq:f} where $w=-t$, namely 
\begin{equation}
 \badat{2}
\int_0^{\infty}\frac{dw}{w}w^{i\frac{\lambda}{2}}\mathcal{B}(rw,-w)&=a \mu^{2\epsilon} \cF_1(r,\epsilon) \int_{0}^{\infty}\frac{dw}{w} w^{i\frac{\lambda}{2}-\epsilon} \,.
\eadat
\end{equation}
The one-loop contribution to the conformally invariant factor is given by
\be
\label{fe}
f_{\rm 1-loop}(r,\bar{r},\epsilon) =  2 a g^2
\mu^{2\epsilon}  \,\cF_1(r,\epsilon) \, \delta(r-\bar{r}) \, \Theta(r-1)\, r^{1+\frac{\Delta}{6}}(r-1)^{\frac{\Delta}{6}} \, \mathcal{I}(\lambda+2i\epsilon)\,.
\ee
While this result shares some similarities with $f_{\rm tree}$ given in~\eqref{eq:f_4gluontree} it notably differs from the tree-level result of the Mellin integral~\eqref{eq:deltalambda} by a shift in the argument. Interestingly, this shift can be re-expressed through the action of the differential operator
\begin{equation}
 \mathcal{I}(\lambda+2i\epsilon)=e^{2i\epsilon \,\partial_\lambda} \mathcal{I}(\lambda)\,.
\end{equation}
This suggests that celestial amplitudes at loop-level may be obtainable from tree-level ones through the action of appropriate differential operators. Indeed, we find that the celestial one-loop amplitude can be written as
\be
\label{eq:Acelestial1LoopOp}
\tilde{\cA}_{\rm 1-loop} = a \,\hat{\cM}^{(1)}_\epsilon \tilde{\cA}_{\rm tree}\,,
\ee
for the {\it celestial one-loop operator}
\be
\label{S1}
\hat{\cM}^{(1)}_\epsilon=\cF_1(r,\epsilon) \hat \cP^\epsilon\,,
\ee
where we defined
\be
\label{eq:pplus}
\hat \cP=\mu^{2}r^{\frac{1}{3}}(r-1)^{\frac{1}{3}}\prod_{i<j} (z_{ij}\bar{z}_{ij})^{-\frac{1}{6}}\,  \exp\left(\frac{i}{2}\sum^4_{k=1} \frac{\partial}{\partial \lambda_k}\right)\,.
\ee
%\be
%\hat{\cS}^{(\epsilon)}=a  \mu^{\Delta-4} \cF_1(r,\epsilon)( [r(r-1)]^{\tfrac{2}{3}}\prod_{i<j} (z_{ij}\bar{z}_{ij})^{-\frac{1}{3}})^{\epsilon/2}\, P(\epsilon)
%\ee
%Thus, we see that the celestial one-loop amplitude in planar $\mathcal{N}=4$ SYM can be recast as an infrared divergent operator $\hat{\cM}^{(1)}_\epsilon$ acting on the celestial tree-level amplitude. 
Notice that $\hat\cP$ is related to the operator $P_{+,k}=e^{\frac{1}{2}(\partial_{h_k}+\partial_{\bar h_k})}$ introduced in~\cite{Stieberger:2018onx} whose  effect is to shift the conformal dimension $\Delta_k \to \Delta_k+1$ of the individual gluons. % or, equivalently, $i\lambda_k \to i\lambda_k+1$. 
There the \emph{sum} over all $P_{+,k}$ was shown to annihilate celestial amplitudes as expected by translation invariance. Here $\hat\cP$ involves instead the \emph{product} over all $P_{+,k}$. Moreover, the prefactor in~\eqref{eq:pplus} can be recast as a correlation function of four scalar primaries with conformal weights $h_{i}=\bar{h}_{i}=\tfrac{1}{4}$. Its appearance is not surprising as the action of the exponential operator shifts all the conformal weights by the same amount $h_i \to h_i-\tfrac{\epsilon}{4}$ and hence the role of the prefactor in~\eqref{eq:pplus} is precisely to cancel these extra factors.

Thus, we see that the celestial one-loop amplitude in planar $\mathcal{N}=4$ SYM can be recast as an infrared divergent operator acting on the celestial tree-level amplitude. The one-loop factor $M^{(1)}_\epsilon$ multiplying the tree-level amplitude in~\eqref{eq:A-1loop} gets promoted to the celestial one-loop operator $\hat \cM^{(1)}_\epsilon$ acting on the celestial tree-level amplitude~\eqref{eq:Acelestial1LoopOp}. In the following we generalize this statement to all loops.

\subsubsection{Celestial gluons at all loops}
\label{ssec:AllLoopSYM}
% 
% We will now show how the operator structure found at one-loop extends to the celestial amplitude at all loops. In the two loops case, the amplitude can be expressed in terms of the one-loop data \cite{Anastasiou:2003kj,Bern:2004cz}  
% \begin{equation}
% M^{(2)}(s,t)= \left(\frac{\mu^2}{-t}\right)^{2\epsilon}\cF_2(r,\epsilon)\,, 
% \end{equation}
% where 
% \begin{equation}
%     \cF_2(r,\epsilon) =\frac{1}{2}\cF_1(r,\epsilon)^2+ f^{(2)}(\epsilon) \cF_1(r,2\epsilon)+C^{(2)}+O(\epsilon)\,,
% \end{equation}
% is the known iterative structure originally found in \cite{Anastasiou:2003kj}. Here $f^{(2)}(\epsilon)$ is a regular function in $\epsilon$, and $C^{(2)}=-\tfrac{1}{2}\zeta_2^2$. This structure appears to persist at higher loops. 

The $\ell$-loop contribution to the four-gluon planar amplitude is 
\begin{equation}
\label{eq:A-Lloop}
A_{\rm \ell-loop}=a^\ell M^{(\ell)}_\epsilon A_{\rm tree}   \,,
\end{equation}
where $M^{(\ell)}_{\epsilon}$ can be written as \cite{Cachazo:2006mq,Spradlin:1900zz}
\be
\label{eq:ll-loop}
M^{(\ell)}_\epsilon =\left(\frac{\mu^4}{s t}\right)^{\frac{\ell\epsilon}{2}}\cG_{\ell}(r,\epsilon)\,.
\ee
For our purposes, using $s=-t/r$, it is more convenient to write this instead as
%where up to $\cO(\epsilon^0)$~\cite{Bern:2005iz}
\be
\label{eq:l-loop}
M^{(\ell)}_\epsilon =\left(\frac{\mu^2}{-t}\right)^{\ell\epsilon}\cF_{\ell}(r,\epsilon)\,,
\ee
where $\cF_\ell =(-r)^{-\frac{\ell \epsilon}{2}} \cG_{\ell}$ has poles starting at $\epsilon^{-2\ell}$. Explicit expression for $\cF_\ell$ to all orders in $\epsilon$ can be obtained in terms of  Mellin-Barnes integral representations for $\ell=2$ in \cite{Smirnov:1999gc}, for $\ell=3$ in \cite{Bern:2005iz} and for $\ell=4$ in \cite{Bern:2006ew}.
Notice that, as in the one-loop case, the dependence of the $\ell$-loop amplitude on the Mandelstam variables is a simple power of $t$ factored out in~\eqref{eq:l-loop} while $\cF_{\ell}(\epsilon,r)$ depends only on the conformal cross ratio, thus making the Mellin integral in~\eqref{eq:f} straightforward to compute. The celestial one-loop expression~\eqref{eq:Acelestial1LoopOp} generalizes to $\ell$-loops as
\be
\label{eq:S-all-loop}
\tilde{\cA}_{\rm \ell-loop}=a^\ell\, \hat{\cM}^{(\ell)}_\epsilon \tilde{\cA}_{\rm tree}\,,
\ee
with the celestial $\ell$-loop operator given by
\be
\label{SL}
\hat{\cM}^{(\ell)}_\epsilon=\cF_{\ell}(r,\epsilon) \hat \cP^{\ell\epsilon}\,.
\ee
We can immediately re-sum the perturbative expansion of the four-gluon celestial amplitude to all loop-orders, yielding   
\be 
\tilde{\cA}_{\rm all\,\,loops}= \hat \cM_\epsilon\, \tilde{\cA}_{\rm tree} \,,
\ee
with
\begin{equation}
\label{eq:AllLoopCelestial}
  \hat \cM_\epsilon= 1+ \sum^{\infty}_{\ell=1} a^\ell \hat{\cM}^{(\ell)}_\epsilon\,.
\end{equation}

Inspired by the BDS exponentiation \cite{Bern:2005iz}, we now show that the infinite sum~\eqref{eq:AllLoopCelestial} can be recast as an exponential operator acting on the celestial tree-level amplitude. 
To do so we make use of the identity
\be
\label{OpID}
1+\sum^{\infty}_{\ell=1} a^\ell \hat\cM^{(\ell)}_\epsilon  = \exp\left(\sum^{\infty}_{L=1}  a^L \left(\hat \cM^{(L)}_\epsilon - \hat X^{(L)}[\hat \cM^{(\ell)}_\epsilon]\right)\right)
\ee
where we introduced
\be
\label{eq:X}
\hat X^{(L)}[\hat \cM^{(\ell)}_\epsilon]= \hat \cM^{(L)}_\epsilon-\left.\log\left(1+\sum^{\infty}_{\ell=1} a^\ell \hat \cM^{(\ell)}_\epsilon \right)\right|_{a^{L} \text{-term}}\,,
\ee
which only depends on the lower-loop operator $\hat \cM^{(\ell)}_\epsilon$ with $\ell <L$.
We obtain
\be
\hat \cM_\epsilon=\exp\left(\sum^{\infty}_{L=1}  a^L \left( \cF_{L}(r,\epsilon) - X^{(L)}[\cF_\ell(r,\epsilon)]\right) \, \hat{\cP}^{L \epsilon}  \right) \,.
\ee
Note that $X^{(L)}$ in the above expression is no longer an operator, as opposed to $\hat X^{(L)}$ in~\eqref{OpID}. At this point, we can make use of the explicit form of $\cF_{\ell}$ known from the BDS formula \cite{Bern:2005iz}
\begin{equation}
\cF_L(r,\epsilon)= X^{(L)}[\cF_\ell(r,\epsilon)]+f^{(L)}_\epsilon \cF_1(r,L\epsilon)+C^{(L)}+\cE_L(r,\epsilon)\,,
\end{equation}
where $\cE_L$ are non-iterating $\cO(\epsilon)$ contributions. With this we find the \emph{celestial} BDS formula
\be
\label{eq:celestialBDS}
\tilde{\cA}_{\rm all\,\,loops} =\exp\left(\sum^{\infty}_{L=1} a^L \left(f^{(L)}_\epsilon \cF_1(r,L\epsilon)+C^{(L)} +\cE_L(r,\epsilon) \right) \, \hat{\cP}^{L \epsilon} \right)  \tilde{\cA}_{\rm tree} \,.
\ee
This demonstrates, for the case of MHV amplitudes in planar $\mathcal{N}=4$ SYM, that celestial amplitudes at loop-level can be obtained through the action of an exponential operator on the celestial tree-level amplitudes.

\section*{Acknowledgements}

We would like to thank Fernando Alday, Guillaume Bossard, Eduardo Casali, Gast\'on Giribet, Andr\'es Gomberoff, Georgios Papathanasiou, Anders Schreiber, and William Torres-Bobadilla for discussions and comments on the draft. The work of H.G. has been supported by FONDECYT grant 11190427. The work of F.R. has been supported by FONDECYT grant 11171148. A.P. was supported in part by the ERC Starting Grant 852386 HoloHair and would like to thank the Facultad de Ingenier\'ia y Ciencias at Universidad Adolfo Ib\'a\~nez for hospitality during her visit.

\appendix
\section{More celestial amplitudes}
\label{app:MoreCelestial}
In this appendix we make use of~\eqref{eq:4point} to analyze tree-level amplitudes in massless QED and finite one-loop Einstein gravity processes.

\subsection{Celestial tree-level massless QED}
\label{app:MoreCelestialQED}
The asymptotic states of massless QED are massless electrons, positrons and photons. We first analyze processes involving only fermions in the external states. The non-vanishing amplitudes correspond to two fermions of positive helicities and two fermions of negative helicities. These are given by (see e.g. \cite{Srednicki:2007qs}) 
\eal{\label{masslessQEDfermions}
A(1^{\half},2^{-\half},3^{-\half},4^{\half})&=-2 e^2 \frac{[14]\langle 23 \rangle}{u} = \frac{2e^2}{r-1} \left(\frac{z_{23} \bar{z}_{14}}{\bar{z}_{23} z_{14}}\right)^{\half}\,,\\
A(1^{\half},2^{-\half},3^{\half},4^{-\half})&=-2 e^2 \frac{[13]\langle 24 \rangle}{t}= \frac{2e^2}{r-1} \left(\frac{z_{24} \bar{z}_{13}}{\bar{z}_{24} z_{13}}\right)^{\half}\,,\\
A(1^{\half},2^{\half},3^{-\half},4^{-\half})&=-2 e^2 [12]\langle 34 \rangle \frac{s}{t u}=\frac{2e^2r^2}{r-1} \left(\frac{z_{34} \bar{z}_{12}}{\bar{z}_{34} z_{12}}\right)^{\half}\,,
}
with the remaining cases given by the complex conjugation. Comparing these expressions with \eqref{eq:4point} yields $\cB(s,t)=2e^2$ for all massless QED amplitudes with external fermions. The conformally invariant expressions for the amplitudes in \eqref{masslessQEDfermions} are given by
\be
f_{\half}(r,\bar{r})= 8\pi e^2  \delta (r-\bar{r})\, \Theta(r-1) \, \delta (\lambda) \times \left\{
  \begin{array}{@{}ll@{}}
    r^{\frac{2}{3}} (r-1)^{\frac{2}{3}}, & (\pm\half,\mp\half,\mp\half,\pm\half)  \\
    r^{\frac{2}{3}} (r-1)^{-\frac{1}{3}}, & (\pm\half,\mp\half,\pm \half,\mp\half)\\
    r^{\frac{8}{3}} (r-1)^{\frac{2}{3}}, & (\pm\half,\pm\half,\mp\half,\mp\half) \end{array}\right.\,.
\ee

Next we consider amplitudes combining two external photons and two external fermions
\begin{align}
\begin{split}
A(1^{\half},2^{-\half},3^{+},4^{-})&=2 e^2 \frac{\langle 24 \rangle^2}{\langle 13 \rangle \langle 23 \rangle} = 2 e^2 (r-1)^{\half} \frac{z_{24}}{\bar{z}_{24}} \left(\frac{\bar{z}_{23}\bar{z}_{13}}{z_{23}z_{13}}\right)^{\half}\,,\\
A(1^{\half},2^{-\half},3^{-},4^{+})&=2 e^2 \frac{\langle 23 \rangle^2}{\langle 14 \rangle \langle 24 \rangle}= 2 e^2 (r-1)^{-\half} \frac{z_{23}}{\bar{z}_{23}} \left(\frac{\bar{z}_{24}\bar{z}_{14}}{z_{24}z_{14}}\right)^{\half}\,\,.
\end{split}
\end{align}
Again we have $\cB(s,t)=2e^2$, and the conformally invariant expression are
\be
f_{\left(1,\half\right)}(r,\bar{r})= 8\pi e^2\,\delta (r-\bar{r})\, \Theta(r-1)  \, \delta (\lambda) r^{\frac{2}{3}} \times \left\{
  \begin{array}{@{}ll@{}}
     (r-1)^{\frac{5}{3}}, & (\half,-\half ,+,-)  \\
     (r-1)^{-\frac{1}{3}}, & (\half,-\half ,-,+)
  \end{array}\right.\,.
\ee 
Hence celestial amplitudes in massless QED are correlators of two-dimensional conformal primaries with spin $1$ and spin $\half$.\footnote{Spin $\half$ conformal wavepackets were recently discussed in~\cite{Fotopoulos:2020bqj} to compute celestial amplitudes in $\mathcal{N}=1$ super Yang-Mills theory.}

\subsection{Celestial pure gravity at one-loop}
\label{app:MoreCelestialgravity}
Finite amplitudes in pure gravity at one-loop are \cite{Bern:1993wt,Dunbar:2010xk} given by
\begin{equation}
\badat{2}
A(1^{++},2^{++},3^{++},4^{++})&=\kappa^4 \frac{s^2-st+t^2}{960}  A(1^{+},2^{+},3^{+},4^{+})^2\,,\\
A(1^{--},2^{++},3^{++},4^{++})&=\kappa^4\frac{s^2-st+t^2}{2880} \left(\frac{st}{u^2}\right)^2 A(1^{-},2^{+},3^{+},4^{+})^2\,,
\eadat
\end{equation}
where $\kappa=\sqrt{32 \pi G_N}$ and we have used the double copy when writing these amplitudes in terms of the four-gluon stripped amplitudes $A(1^{\pm},2^{+},3^{+},4^{+})$ given in~\eqref{eq:12} and~\eqref{eq:14}. For the corresponding celestial amplitudes, we find
\be
f_{\rm 1-loop}(r,\bar{r})= \frac{\kappa^4}{240}\,  \cI (\lambda-2i)\,  \delta (r-\bar{r})  \,\Theta(r-1) \times \left\{
  \begin{array}{@{}ll@{}}
    r^2+r+1, & (++,++,++,++)  \\
    \frac{r^2+r+1}{3(r-1)^2}, & (--,++,++,++)
  \end{array}\right.\,,
\ee
with the Mellin integral $\cI(\lambda)$ defined in~\eqref{eq:deltalambda}.
Note that this is the same type of divergence observed in celestial tree-level graviton amplitudes~\cite{Stieberger:2018edy}.\footnote{A regularization of classical celestial graviton amplitudes is given by an imaginary shift of $\lambda$ corresponding to the insertion of gravitons of integer conformal dimension such as the stress tensor or its shadow~\cite{Puhm:2019zbl}.} In string theory, these are rendered finite due to the exponentially damped behavior of the amplitudes at high energies~\cite{Stieberger:2018edy}. Since the soft UV behavior holds perturbatively to all orders in the string loop expansion for both closed~\cite{Gross:1987kza,Gross:1987ar} and open strings~\cite{Gross:1989ge}), one would expect convergence of the Mellin integrals of the corresponding celestial string amplitudes. In the $r\to \infty$ limit one should then recover the above celestial one-loop gravity amplitude.

\providecommand{\href}[2]{#2}\begingroup\raggedright\endgroup

\end{document}